\documentclass{sig-alternate-05-2015}

\usepackage[ruled,vlined]{algorithm2e}

    \newcommand\BlockIf[1]{\KwSty{Start If} \\ #1 \\ \KwSty{End If}}
    
    \newcommand\BlockElse[1]{\KwSty{Start Else} \\ #1 \\ \KwSty{End Else}}

\begin{document}

\setcopyright{acmcopyright}

\doi{10.475/123_4}

\isbn{123-4567-24-567/08/06}

\conferenceinfo{RECSYS}{2016 Boston, Massachusetts USA}

\acmPrice{\$15.00}

%
\conferenceinfo{RECSYS}{'16 Boston, Massachusetts USA}

\title{Feedback-based Approach to Introduce Freshness in Recommendations}

%
%
%
%
%

\numberofauthors{2} 
%
\author{
%
%
\alignauthor
Hari Krishna Malladi\\
       \affaddr{Gwynniebee India Pvt. Ltd.}\\
       \affaddr{Bangalore, India}\\
       \email{hari.krishna@gwynniebee.com}
 2nd. author
\alignauthor
Saikiran Thunuguntla\\
       \affaddr{Gwynniebee India Pvt. Ltd.}\\
       \affaddr{Bangalore, India}\\
       \email{skiran@gwynniebee.com}
}


\maketitle
\begin{abstract}
Recommender systems usually face the problem of serving the same recommendations across multiple sessions regardless of whether the user is interested in them or not, thereby reducing their effectiveness. To add freshness to the recommended products, we introduce a feedback loop where the set of recommended products in the current session depend on the user's interaction with the previously recommended sets. We also describe ways of addressing freshness when there is little or even no direct user interaction. We define a metric to quantify freshness by reducing the problem to measuring temporal diversity.
\end{abstract}

%
%


%
%

%
%



\section{Introduction}

Recommender systems have a variety of applications today, such as ecommerce, financial services, social networks, news and the like. They find applications in places where there is a necessity to personalize the information to be presented to each user out of a much larger and varied collection. A persistent problem in recommender systems is that of freshness. Subsequent calls to the recommender from the same user return similar results. The user tends to lose interest in the products recommended and the system becomes ineffective. This problem has been addressed in various ways~\cite{netflix,youtube} to meet specific requirements. This paper attempts to introduce a metric to quantify freshness and proposes two approaches to introduce freshness into a recommender system as a postprocessing step in an ecommerce setting.~\cite{cqafreshness} deals with a similar problem, but related to a Community Question Answer (CQA) use case.~\cite{codereporec} deals with computing fresh recommendations for programs similar to the program being edited from a code repository.

Feedback loops are not uncommon in recommendations and work has been done on using feedback loops to add diversity to recommendations~\cite{feedback}. We introduce a feedback loop based mechanism which ensures that only the recommendations that the user is potentially interested in are shown. It decays with time and a product that has been marked as potentially uninteresting might be recommended again in the future unless a recommendation of higher relevance has taken its place. We incorporate time decay as the user would probably be interested in the product in the future. This approach would require instrumentation of the recommender system to monitor the clicking behaviour of the user and the user's dwell time. In the absence of any data to estimate the user's interest, we shuffle the recommendations each time they are presented to the user, in such a way that there are a reasonably high number of combinations and closely related products are not clustered together. 

We define a metric to quantify freshness by reducing the problem of measuring freshness to that of measuring diversity in the time domain. The metric measures the unique recommendations that have been served across a sliding window of time. We also describe an algorithm to use the metric to add freshness using a feedback loop.

The paper is structured as follows. Section 2 discusses about recommenders where freshness is an inherent feature. It deals with the cases where the recommendations are indeed fresh and cases where they are not. Section 3 deals with the proposed solutions, pseudocode and reasoning behind them. Section 4 defines the freshness metric. Section 5 concludes the paper with mentions about the experiments being performed to validate the ideas in the paper.

\section{Inherent Freshness}

The presence of freshness depends upon the type of recommender used and the application where it is used in. Some applications, with some types of recommenders, have an inherent freshness factor to them, while others, such as YouTube, can introduce freshness by simply retrieving a large number of recommendations and displaying them batch wise~\cite{youtube}. This works because, in the YouTube setting, the number of videos that can be recommended with a reasonable relevance is enormous. However, the same cannot be said about smaller ecommerce companies where the inventory is much smaller and changes much less often.

Freshness is inherent in applications where the inventory is large and is fast changing. New items are constantly added to the inventory freqently and thus make their way into recommendations. In applications where the user base is fast changing and a collaborative filter is used, freshness could be inherent as the user behaviours differ from person to person and new users bring in new behavious, thus possibly resulting in diverse recommendations. Real-time recommender systems dealing with streaming data show inherently fresh recommendations~\cite{streamrec}.

Freshness is not an inherent characteristic of recommender systems which use a content based filter and have a slow changing product base. The only way that freshness can be inherent in these systems is if users change their behaviour quickly, allowing the recommender to find more diverse products which would relate to the user's preferences at a given time.

\section{Proposed Algorithm}

The implementation details of the proposed algorithm tend to differ from application to application as the requirements of the recommender changes and so does the instrumentation of events. We elaborate on one particular use case. Our use case deals with a subscription-based ecommerce application, where customers don't have to spend money for every transaction that they perform. Instead, they spend money on a monthly or yearly rental basis. In a $k$-product plan, $k$ products from the user's selected priority list are shipped to his/her location. The user can browse the product page of the website and add any number of products to his/her priority list at no additional cost. Once a shipped product is returned, another product is shipped to him/her to replace it. At any given point in time, $k$ products exist at the user's end.

\subsection{Feedback Loop}

The algorithm that we propose for freshness uses a feedback loop, which uses instrumentation in the recommender system to estimate which products the user is probably not interested in at the moment. We use the following events instrumented in our recommender for analysis.

\begin{itemize}
 \item Recommendations served. 
 \item Recommendations viewed.
 \item Recommendations clicked.
 \item Products purchased from the ones recommended.
 \item Dwell time after a recommendation has been clicked.
\end{itemize}

The recommender uses an array called NegativeWeights. This array has one node per product in the inventory. The higher the value of the node corresponding to a product, the higher the user's disinterest (estimated) in that particular product. The feedback loop increments these values based on user actions. This array is used to calculate the final relevance score while serving the recommendations. The products with high NegativeWeights score could end up at the end of the list of recommended products or might not be recommended at all. However, if a particular product's relevance score far exceeds the scores of all other products in the recommended set, it might still show high up on the recommended products list until further user action decreases its relevance to a point where it is no longer recommended. The exact increments of NegativeWeights is subjective to the application that it's being used in.

For our use case, we hypothesize that a user is not interested in a product if he clicks it and does not add it to his/her priority list. Our hypothesis that the user is not interested is strengthened if the dwell time is high. This could mean that the user has done some survey on the product such as reading reviews and looking at its ratings and is quite informed about it. Thus, we increment the NegativeWeights value for that product accordingly. This approach is justified by the fact that in our use case, the user spends no additional money to add a product to his/her priority list. So, the dwell time would be used in actually researching about the product and not on its price.

We propose a factor of time decay for the NegativeWeights array. This ensures that relevant recommendations get a chance to return to the list of recommendations served after some time. This could be implemented in several ways.

\begin{itemize}
 \item Reinitialize the value of a node in the NegativeWeights array to 1 if that value is older than \textit{n} days.
  \item Reinitialize the value of a node in the NegativeWeights array to 1 if that product has been chosen for recommendation but could not be recommended due to its NegativeWeights value more than \textit{k} times.
 \item Reinitialize the entire NegativeWeights array to 1 after \textit{k} recommendations have been served.
 \item Reinitialize the entire NegativeWeights array to 1 after \textit{n} days.
\end{itemize}

The values of \textit{k} and \textit{n} depend on the application and vary from case to case. In some applications, such as TV show recommenders, time decay itself might not be relevant. Our use case uses the number of recommendations served instead of the actual time to reinitialize the NegativeWeights array. The feedback loop algorithm is detailed in Algorithm 1. The function $f$, which is used to add weights to the NegativeWeights array has to be determined. We chose a linear function such as $f(\bar{X}) = k \cdot \bar{X}$ for our implementation. The value of $k$ is to be decided experimentally.


\begin{algorithm}[h]
\KwIn{Inventory, User data, $MaxDecay$}
\KwOut{$RecProdList$}

\nl Initialize $NegativeWeights$ to $1$\;
\nl Generate the array, $RelevanceScores$, which includes the relevance score of every product in the inventory\;
\nl $FinalScores = RelevanceScores - NegativeWeights$\;
\nl Serve recommendations with the highest $FinalScores$ as $RecProdList$\;
\nl Record click behaviour and dwell time and create the array, $DwellArray$, containing dwell times on all products\;
\nl NegativeWeights = NegativeWeights + $f(DwellArray)$\;
\nl For each $i$, $NegativeWeights[i] = 1$ if $NegativeWeights[i]$ is older than $MaxDecay$ days. This can be achieved by tracking it with a $MaxDecay$ array.\;
\nl Goto Step 2\;
    \caption{{\bf Freshness Through Feedback Loop} \label{Algorithm}}

\end{algorithm}

\subsection{Shuffle Logic}

Shuffling helps add an element of diversity to a set of recommendations when there is no event information to be used. However, shuffling would also mean that the user might not always be served recommendations in the descending order of relevance. We highlight two broad approaches to shuffling.

\begin{itemize}
 \item Shuffle the entire recommendation set.
 \item Shuffle the recommendations batch-wise, after partitioning the recommended set into batches based on relevance scores. 
\end{itemize}

The first approach would result in a total mix of relevance scores, but yield far higher combinations than the second approach. The second approach maintains the ordering with respect to the relevance scores up to the granularity of the batches, unlike the first approach. The first approach will result in at most $n!$ combinations, given $n$ recommendations, whereas the second one would result in $(\frac{n}{h})! \times h$ combinations, where $n$ recommendations are split into blocks of $h$ recommendations each.

The shuffle approach should be used only when freshness has a higher priority as compared to the order in which recommendations are served. In our use case we ensure that, wherever possible, no two products of the same brand are placed next to each other while the recommendations are being served. We use the first approach mentioned above, while maintaining this property.

The shuffle algorithm is detailed in Algorithm 2. 

\begin{algorithm}[h]
\KwIn{$RecList$, $p$}
\KwOut{$ShuffList$}

\nl Partition $RecList$ into blocks of length $p$ each, resulting in $m$ partitions, $R_1,...,R_m$\;
\nl Arrange each $R_i$ such that products with the same brands do not end up next to each other, wherever possible\;
\nl Swap a product of brand $b$ in $R_i$ with another product of the same brand from $R_1,...,R_{i-1},R_{i+1},...,R_m$, chosen at random\;
\nl Perform step 3 for all brands in every partition\;
\nl Output the final list as $ShuffList$\;
    \caption{{\bf Freshness Through Shuffling} \label{Algorithm}}
\end{algorithm}

 An alternate approach to shuffling could be to use the following metric as a feedback mechanism, instead of instrumentation. However, the actual effectiveness of this feedback mechanism as compared to shuffling depends on the application it's being used in.

\section{Freshness Metric}

We define a metric to quantify freshness in this section. Freshness can be thought of as diversity in the time domain. This is in contrast to the intra-list diversity~\cite{ild}, enhanching which is part of most recommender systems. Our freshness metric is based on the metrics defined in~\cite{temporal}. We use a modified version of the \textit{top-N novelty} metric defined in Section 3 of~\cite{temporal}. The freshness metric applies by default to all the $t$ recommendations served. The \textit{top-N novelty} metric is modified to incorporate a time-decay factor as below.

$$Freshness = \frac{|R \setminus A_k|}{t},$$ where $A_k$ is the set of all products recommended in the past $k$ calls to the recommender. This metric is oblivious to shuffling and it has a positive value only when a unique product is recommended. The feedback loop in the previous section can be modified to feed the metric back, instead of the instrumented events data. This could be done in cases where there are no events, as a (possibly better) alternative to shuffling. Also, this can be merged with the events data to create a feedback loop which works with not just the instrumented events but also the final outcomes. 

A metric-based feedback loop approach is detailed in Algorithm 3.

\begin{algorithm}[h]
\KwIn{Inventory, User data, $MaxDecay$, $FreshnessThreshold$, $ProdRecTillNow$, $count$}
\KwOut{$RecProdList$, $ProdRecTillNow$, $count$}

\nl Obtain a list of recommendations, $RecProdList$, by calling the recommendations API. $RecProdList$ has $t$ products\;
\nl Calculate $$Freshness = \frac{|RecProdList \setminus ProdRecTillNow|}{t};$$
\nl If $Freshness \leq FreshnessThreshold$, replace a product in $RecProdList$ with the next most relevant one\;
\nl Repeat Step 3 till $Freshness \geq FreshnessThreshold$\;
\nl \uIf{ $count < MaxDecay$ }{\BlockIf{ $ProdRecTillNow = ProdRecTillNow \cup RecProdList;~  count++\;$}}
    \Else{\BlockElse{$ProdRecTillNow = \phi;~ count = 0\;$}}
\nl Return $RecProdList$, $ProdRecTillNow$ and $count$\;

\caption{{\bf Feedback Loop with Freshness Metric} \label{Algorithm}}

\end{algorithm}

\section{Conclusions}
This is a work in progress and the experiments related to it are being carried out. We are evaluating the effectiveness of the feedback loop mechanism using a content and a collaborative filter through A/B testing. This will also help build our confidence in the defined metric. The algorithm and metric might require revision if the experiments and testing suggest so. 

\bibliographystyle{abbrv}
\bibliography{sigproc}  
%
%

\end{document}